\documentclass{moriond}

\def\be{\begin{equation}}
\def\ee{\end{equation}}
\def\bea{\begin{eqnarray}}
\def\eea{\end{eqnarray}}

\begin{document}
\vspace*{4cm}
\title{MEASUREMENT OF BARYON ELECTROMAGNETIC FORM FACTORS AT BESIII}

\author{ C. MORALES MORALES on behalf of the BESIII Collaboration}

\address{Helmholtz-Institut Mainz, 55099 Mainz, Germany}

\maketitle\abstracts{
The Beijing $e^+e^-$-collider (BEPCII) is a double-ring symmetric collider running at center-of-mass energies  between 2.0 and 4.6 GeV. This energy range allows the BESIII-experiment to measure baryon electromagnetic form factors in direct $e^+e^-$-annihilation and in initial state radiation processes. In this paper, results on $e^+e^-\rightarrow p\bar{p}$ and $e^+e^-\rightarrow \Lambda \bar{\Lambda}$ based on data collected by BESIII in 2011 and 2012 are presented. Expectations from the BESIII high luminosity energy scan from 2015 and from radiative return at different center-of-mass energies are also reported.}

\section{Introduction}

Form factors (FFs) account for the non point-like structure of hadrons. Depending on its spin, s, a hadron has 2s+1 form factors. The FFs are analytic functions of the momentum transferred by the virtual photon, $q$. They are real in the space-like region ($q^2<0$) and complex in the time-like region ($q^2>0$) for $q^2 > 4m_\pi^2$. FFs at $q^2<0$ are determined by elastic scattering of electrons from hadrons available as targets. FFs at  $q^2>0$ are measured in annihilation  processes \mbox{$e^+e^- \leftrightarrow h \overline{h}$}.

The Born differential cross section of the $e^+e^-$-annihilation into a baryon-antibaryon pair, \mbox{$e^+e^- \rightarrow B\overline{B}$}, in $e^+e^-$ center-of-mass (c.m.) reads~\cite{Zichichi} 
\begin{equation}
\label{diff}
\frac{d\sigma^{\mathrm{Born}}(q^2,\theta_{B}^*)}{d\Omega} = \frac{\alpha^2\beta C}{4q^2} \left [(1 + \mathrm{cos}^2\theta_{B}^*) |G_M (q^2)|^2 +  \frac{1}{\tau}\mathrm{sin}^2\theta_{B}^*|G_E(q^2)|^2 \right ],
\end{equation}
with $G_E$ and $G_M$ the Sachs FFs, $\theta_{B}^*$ the polar angle of the baryon, $m$ the baryon mass, $\tau = 4m^2/q^2$ and $\beta = \sqrt{1-1/\tau}$. The Coulomb factor,  $C = y/(1-\mathrm{exp}(-y))$ with $y = \pi \alpha / \beta$, accounts for the electromagnetic $B \overline{B}$ interactions of point-like baryons \cite{Tzara}. Angular integration of the previous equation gives the total cross section:
\begin{equation}
\sigma^{\mathrm{Born}}(q^2) = \frac{4\pi\alpha^2\beta C}{3q^2} \left[ |G_M(q^2)|^2 + \frac{1}{2\tau} |G_E(q^2)|^2 \right ]. 
\label{totalcs}
\end{equation}
An effective form factor (EFF) can be defined as 
\begin{equation}
|G(q^2)|^2 =  \frac{2\tau |G_M(q^2)|^2 + |G_E(q^2)|^2}{2\tau +1} = \frac{\sigma^{\mathrm{Born}}(q^2) }{(1+\frac{1}{2\tau})(\frac{4\pi\alpha^2\beta C}{3q^2})} , 
\label{effectiveff}
\end{equation}
which is equivalento to $|G_M(q^2)|$ under the working hypothesis $|G_E(q^2)| = |G_M(q^2)|$. The simultaneous extraction of $|G_E|$ and $|G_M|$ without any assumption is only possible by measuring the angular distributions of the outgoing particles (Eq.~\ref{diff}).

The process of $e^+e^-$-annihilation can also be accompanied by the emission of one or several high energy photons from the initial state (ISR).
The differential cross section reads:

\begin{equation}
\frac{d^2\sigma^{\mathrm{ISR}}} {dq^2d\theta_{\gamma}^*} = \frac{1}{s} \cdot W(s,x,\theta_{\gamma}^*) \cdot \sigma^{\mathrm{Born}}(q^2),
\label{differential}
\end{equation}
where $x = 2E_\gamma^*/\sqrt{s} = 1 - q^2/s$, $\sqrt{s}$ is the c.m. energy of the collider, and $E_\gamma^*$ and $\theta_\gamma^*$ are the energy and polar angle of the ISR photon in the c.m. The radiator function $W(s,x,\theta_{\gamma}^*)$ describes the probability of the ISR photon emission~\cite{Radiator}.
Due to the ISR photon emission, the hadronic invariant mass in the final state is reduced below $\sqrt{s}$ up to the production threshold of the hadronic state.

\section{The BESIII experiment and data sets}
BEPCII is a double ring $e^+e^-$ symmetric collider running at $\sqrt{s}$ from 2.0 to 4.6 GeV. The design luminosity is $1 \times 10^{33}$~$\mathrm{cm}^{-2}\mathrm{s}^{-1}$ at a beam energy of 1.89 GeV. BESIII is a cylindrical detector which covers 93\% of the full solid angle~\cite{bes3}. It consists of the following sub-detectors: a Multilayer Drift Chamber (MDC);
a Time-of-Flight plastic scintillator (TOF); 
a CsI(Tl) Electro-Magnetic Calorimeter (EMC);
a superconducting magnet of 1T
and  a Muon Chamber (MUC).
BESIII has accumulated the world$^\prime$s largest samples of $e^+e^−$-collisions in the tau-charm region. Furthermore, in 2015 BESIII peformed a high luminosity scan in 21 energy points between $\sqrt{s}=$ 2.0 and 3.08 GeV, with about 555  $\mathrm{pb^{-1}}$ luminosity. These statistics are the highest in this energy region.
\label{subsec:prod}

\begin{figure}[t!]
\begin{minipage}{0.34\linewidth}
\centerline{\includegraphics[width=1\linewidth]{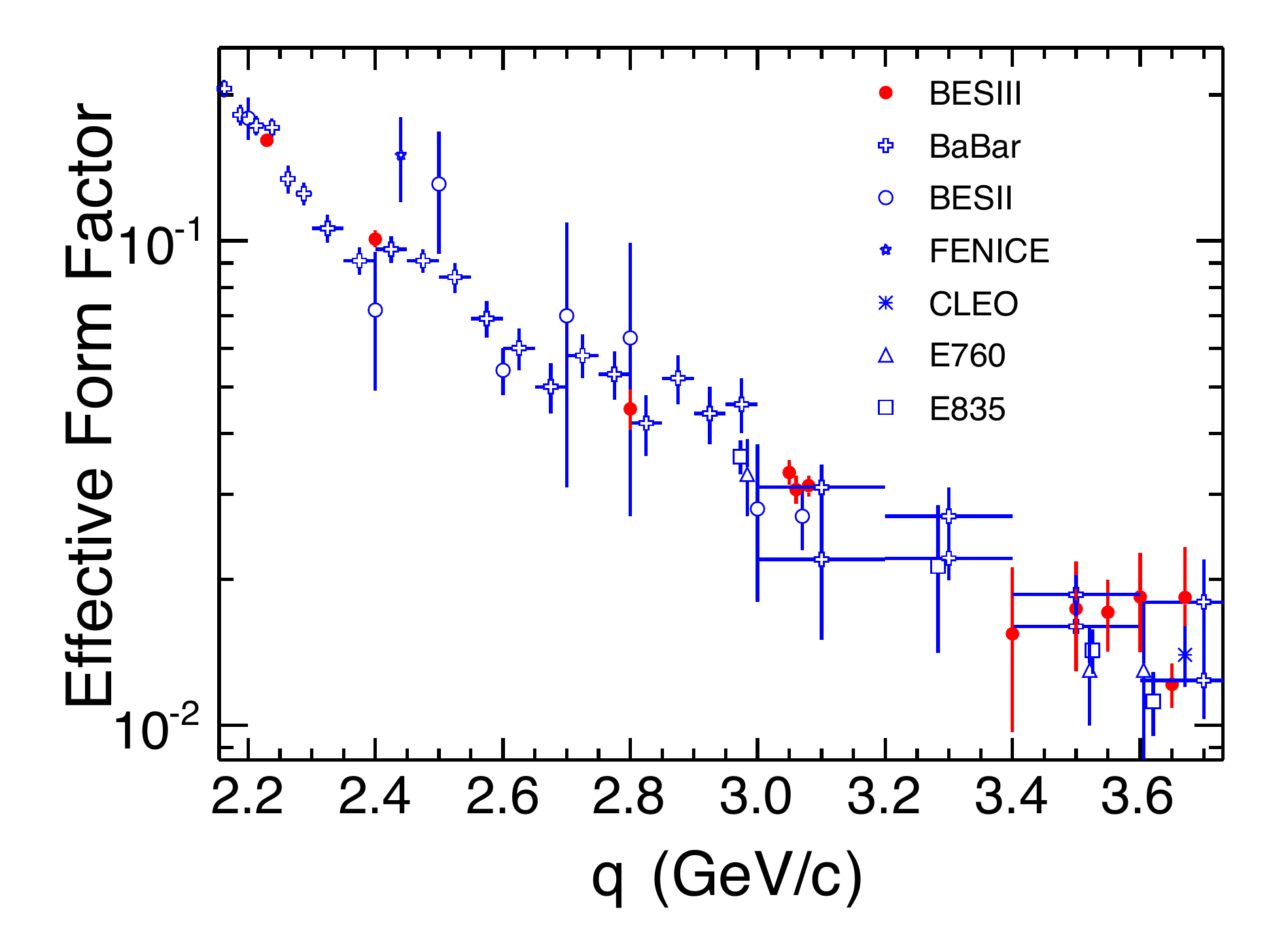}}
\end{minipage}
\hfill
\begin{minipage}{0.32\linewidth}
\centerline{\includegraphics[width=1\linewidth]{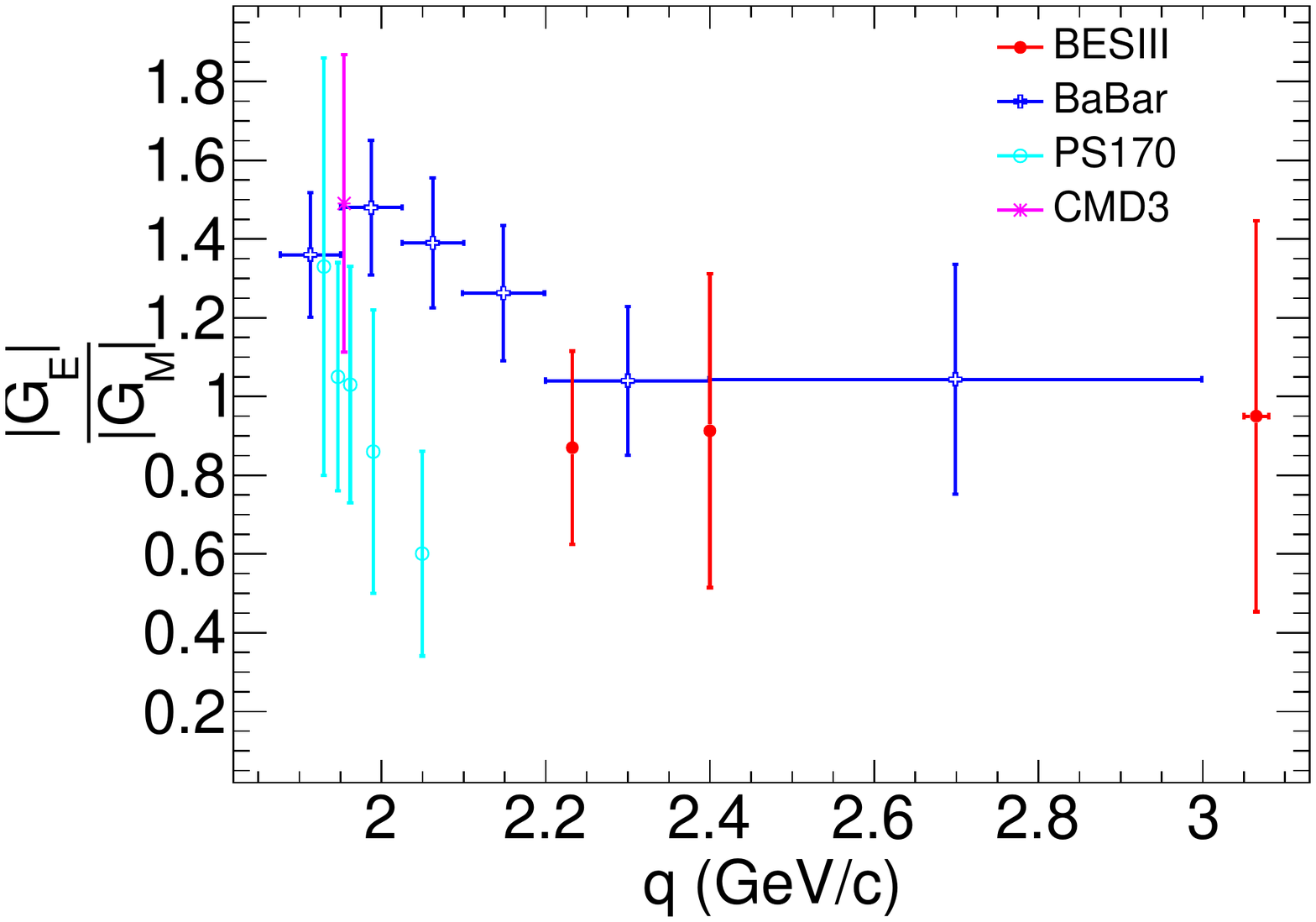}}
\end{minipage}
\hfill
\begin{minipage}{0.32\linewidth}
\centerline{\includegraphics[width=1\linewidth,height=0.7\linewidth]{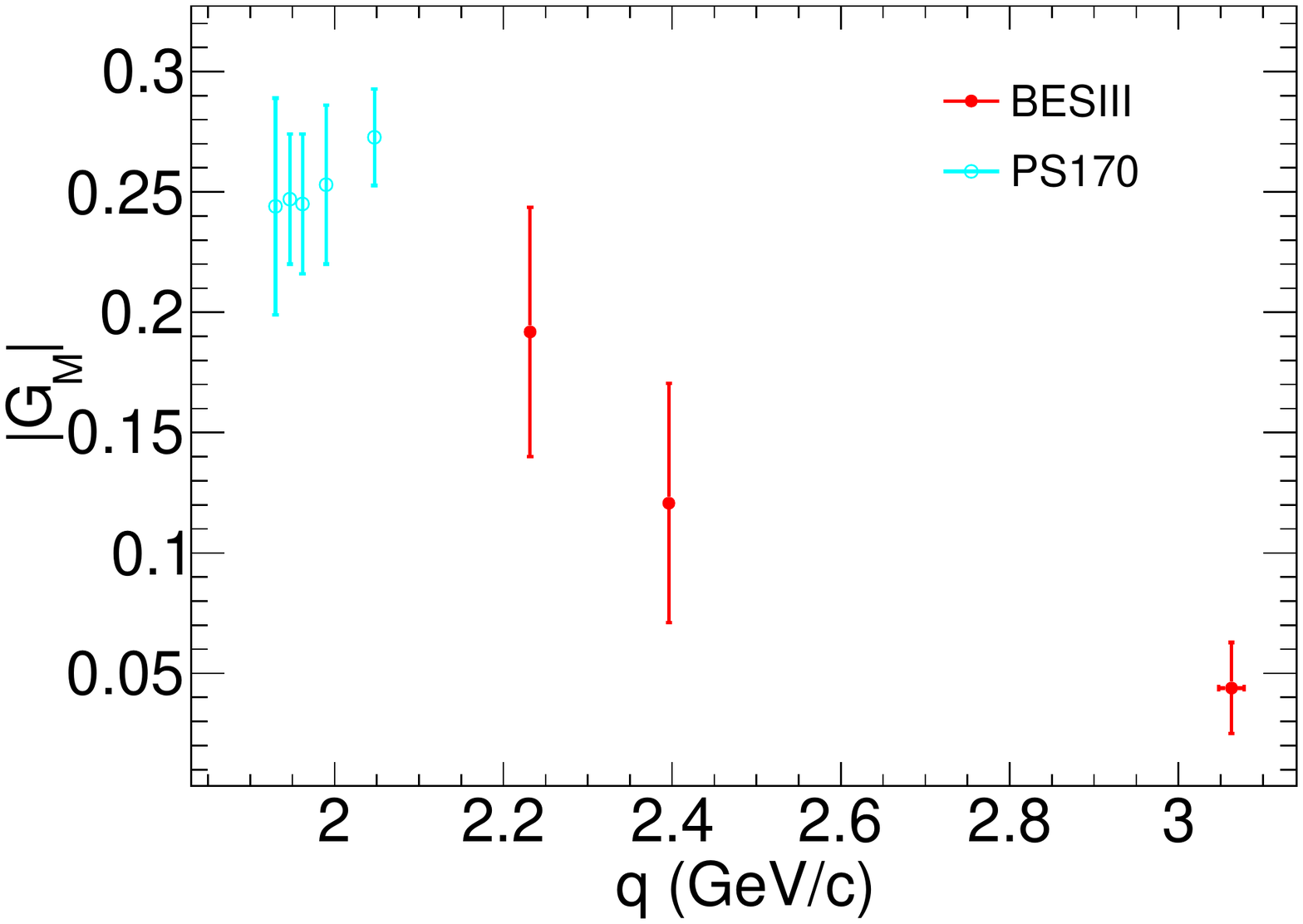}}
\end{minipage}
\caption[]{(Left) Proton effective form factor, (center) $|G_E/G_M|$ and (right) $|G_M|$ in the time-like region.}
\label{FFxiaorong}
\end{figure}

\section{Measurement of $e^+e^- \rightarrow p \overline{p}$}

In June 2015, BESIII published the measurement of the channel $e^+e^-\rightarrow p\bar{p}$ at 12 c.m. energies between 2.2324 and 3.6710 GeV~\cite{xiaorong}. These data were collected in 2011 and 2012 and correspond to a luminosity of 157 $\mathrm{pb^{-1}}$. The Born cross section was extracted according to 
\begin{equation}
\sigma_\mathrm{Born} =  \frac{N_{obs} - N_{bkg}}{\mathcal{L}\cdot\epsilon(1+\delta)}, 
\label{xsmeasured}
\end{equation}
where the number of background events is subtracted from the observed signal event candidates, normalized with the luminosity at each scan point, $\mathcal{L}$, and corrected with the selection efficiency of the process, $\epsilon$,  times the ISR radiative correction factor up to next-to-leading order (NLO), $(1+\delta)$. The ConExc generator~\cite{conexc} was used both for the efficiency and the radiative factor evaluation. 
The accuracy in the cross section measurements was between 6.0$\%$ and 18.9$\%$ up to $\sqrt{s} < 3.08$ GeV. The EFF was extracted according to Eq.~\ref{effectiveff} and is shown in Fig.~\ref{FFxiaorong} (left) together with previous experimental results~\cite{bes,CLEO,BABARpp,PS170,E760,E835,cmd3}. A fit to the polar angular distribution of the proton in c.m. was performed according to Eq.~\ref{diff} and $|G_E/G_M|$ and $|G_M|$ were extracted for three energy points (Fig.~\ref{FFxiaorong} (center and right)). 
While the measurements by the different experiments concerning the EFF show very good agreement, this is not the case of $|G_E/G_M|$, where the measurements by BaBar~\cite{BABARpp} and PS170~\cite{PS170} disagree for low $q^2$.

\section{Measurement of $e^+e^- \rightarrow \Lambda \overline{\Lambda}$}
\begin{figure}[t!]
\begin{center}
\begin{minipage}{0.33\linewidth}
\centerline{\includegraphics[width=1\linewidth]{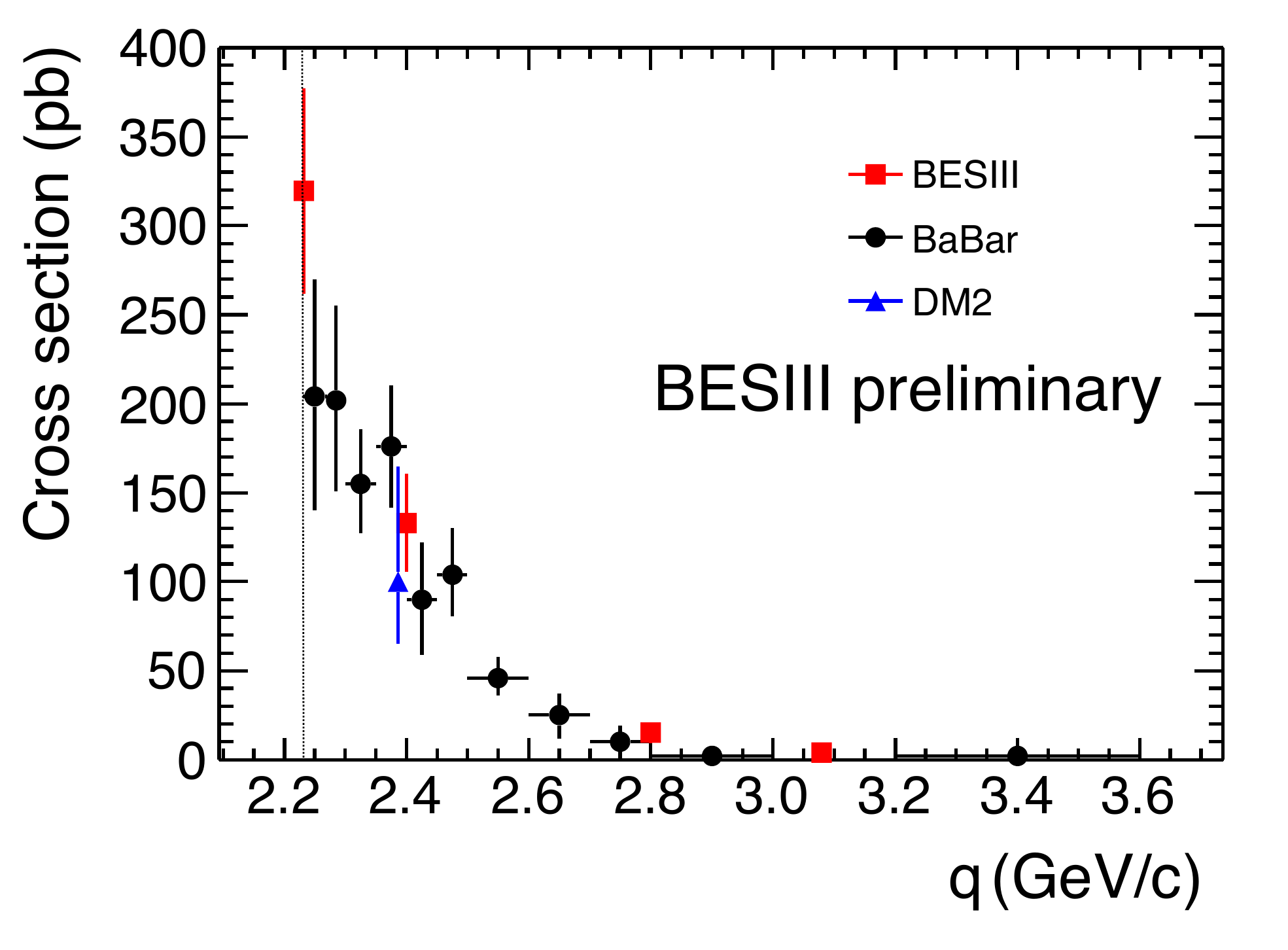}}
\end{minipage}
\begin{minipage}{0.33\linewidth}
\centerline{\includegraphics[width=1\linewidth]{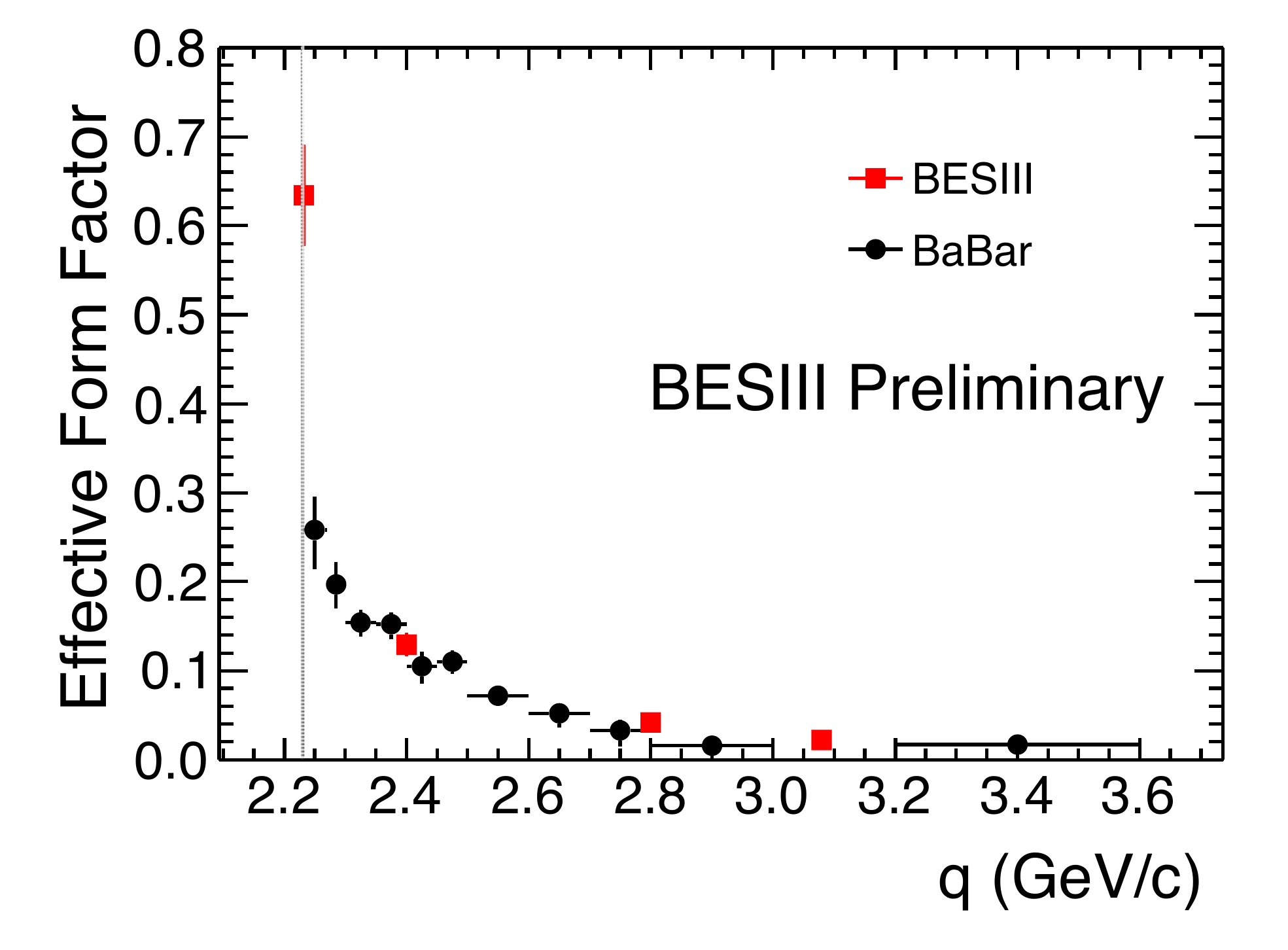}}
\end{minipage}
\end{center}
\caption[]{(Left) Measurements of $e^+e^{-} \rightarrow \Lambda \bar{\Lambda}$ cross section and (right) $\Lambda$ effective form factor.}
\label{lambda}
\end{figure}

BESIII has preliminary results on the measurement of the channel $e^+e^{-} \rightarrow \Lambda \bar{\Lambda}$. The analysis is based on $40.5~\mathrm{pb}^{-1}$ collected in 4 different scan points during 2011 and 2012. The lowest energy point is at 2.2324 GeV, only 1 MeV above the $\Lambda\bar{\Lambda}$-threshold. This makes it possible to measure the cross section almost at threshold. To use as much statistics as possible, both events where $\Lambda$ and $\bar{\Lambda}$ decay to the charged mode ($\mathrm{BR}(\Lambda \rightarrow p\pi^-) = 64\%$) and events where the $\bar{\Lambda}$ decays to the neutral mode  ($\mathrm{BR}(\bar{\Lambda} \rightarrow \bar{n} \pi^0) = 36\%$) are selected. In the first case, the identification relies on finding two monoenergetic charged pions and a possible $\bar{p}$-annihilation.
In the second case, the $\bar{n}$-annihilation is identified through the use of Multivariate Analysis of EMC variables. Additonally, a monoenergetic $\pi^0$ is reconstructed to fully identify the channel. 
For the higher energy points, only the charged decay modes of $\Lambda$ and $\bar{\Lambda}$ are reconstructed by identifying all the charged tracks and using the event kinematics. The preliminary results on the measurement of the Born cross section are shown in Figure~\ref{lambda} (left) together with previous measurements~\cite{LAMBDAS,LAMBDAS2}. The cross section at threshold is found to be $318 \pm 47 \pm 37$ pb. Given that the Coulomb factor in Eq.~\ref{diff} is equal to 1 for neutral baryon pairs, the cross section is expected to go to zero at threshold. This result confirms BaBar$^\prime$s measurement~\cite{LAMBDAS2} but with much higher $q^2$ accuracy. The BESIII measurement improves at least by 10$\%$ previous results at low $q^2$ and even more above 2.4 GeV. The lambda EFF extracted from the cross section measurement is shown in Fig.~\ref{lambda} (right).

\section{Summary and outlook}

Using data samples colllected in test runs in 2011 and 2012, BESIII has measured the Born cross section of $e^+e^- \rightarrow p \overline{p}$ at 12 c.m energies from 2.2324 to 3.6710 GeV. The corresponding effective electromagnetic form factor was also extracted. In addition, the ratio of electric to magnetic form factors, $|G_E/G_M|$, and $|G_M|$ were extracted for the three data samples with larger statistics ($\sqrt{s} =$ 2232.4, 2400.0 and 3050.0-3080.0 MeV). The measured cross sections were in agreement with recent results from BaBar, improving the overall uncertainty by about $30\%$. The $|G_E/G_M|$ ratios were close to unity and consistent with BaBar results in the same $q^2$ region. At present, the precision in the measurement of $|G_E/G_M|$ is dominated by statistics. Using the high luminosity $e^+e^{-}$ scan data collected by BESIII in 2015, statistical accuracies below $10\%$ are expected to be achieved. During this energy scan, BESIII collected the world's largest luminosity in c.m. energies between 2.0 and 3.08 GeV. In addition, using the $7.4~\mathrm{fb}^{-1}$ data collected at resonances above 3.77 GeV BESIII can also measure the $e^+e^- \rightarrow p \overline{p}$ channel using radiative return. The ISR visible luminosity for this channel is comparable to BaBar$^\prime$s one using their total collected luminosity ($\sim$ 500$~\mathrm{fb^{-1}}$). Thus, the analysis of this channel will also lead to very competitive results.

Measurements with unprecedented statistics are also expected for the channels $e^+e^{-} \rightarrow n \bar{n}$ and $e^+e^{-} \rightarrow n \bar{n} \gamma$. This will allow the determination of the $e^+e^{-} \rightarrow n \bar{n}$ Born cross section and the neutron effective form factor with much higher accuracies than the ones achieved so far~\cite{dm1,dm2,fenice,snd} and in a much larger $q^2$ region. Furthermore, the first measurement of the $|G_E/G_M|$ of the neutron in the time-like region will be possible for serveral $q^2$.

The process $e^+e^-\rightarrow \Lambda \overline{\Lambda}$ has also been studied using data samples at $\sqrt{s}$ = 2.2324, 2.400, 2.800 and 3.080 GeV. The cross section has been measured for the first time very close to threshold and found to be $318 \pm 47 \pm 37$ pb
 at $\sqrt{s}$ = 2.2324 GeV. The substantial cross section 1.0 MeV above threshold is unexpected for neutral baryon pairs and points towards a more complicated underlying physics scenario. The Born cross sections at other energies have been measured and found to be consistent with previous experimental results, but with improved precision. Besides, the corresponding electromagnetic effective form factor of $\Lambda$ was deduced. Using BESIII scan data from 2015, a full determination of the lambda FFs is possible. The imaginary part of the FFs leads to polarization observables that can be observed by studying $\Lambda$ parity violating decays. This is the case of the relative phase between the lambda FFs. The expected statistical accuracies in the measurement of the lambda polarization range between 6 and 17$\%$. The corresponding accuracies for the $|G_E/G_M|$ of the lambda range between 14 and 29$\%$. Similar measurements might also be possible in other hyperon channels like $e^+e^- \rightarrow \Lambda \bar{\Sigma}^0, \Sigma^0\bar{\Sigma}^0, \Sigma^+\bar{\Sigma}^-, \Xi^0\bar{\Xi}^0, \Sigma^-\bar{\Sigma}^+, \Omega^-\bar{\Omega}^+,\Lambda^+_c \bar{\Lambda}^-_c $.

\section*{Acknowledgments}
This work was supported  by the German Research Foundation DFG under the Collaborative Research Center CRC-1044.

\section*{References}


\end{document}